\documentclass[11pt]{article}
\usepackage{moriond,epsfig,wasysym}

\def\Journal#1#2#3#4{{#1} {\bf #2}, #3 (#4)}

\def\NIMPRA{{\em Nucl. Instrum. Methods Phys. Res.} A}
\def\NPB{{\em Nucl. Phys.} B}

\def\EPJC{{\em Eur. Phys. J.} C}

\def\be{\begin{equation}}
\def\ee{\end{equation}}
\def\bea{\begin{eqnarray}}
\def\eea{\end{eqnarray}}

\begin{document}
\vspace*{4cm}
\title{SEARCHES FOR BSM HIGGS AT THE TEVATRON}

\author{ L. SCODELLARO\\ (on behalf of the CDF and D$\O$ Collaborations)}

\address{Instituto de Fisica de Cantabria, Avda de los Castros s/n,\\
Santander 39005, Spain}

\maketitle\abstracts{
In this paper, we present the latest results of the searches for beyond standard 
model Higgs boson production at the Tevatron collider of Fermilab. Analyses have been
carried out on samples of about 1-4~fb$^{-1}$ of data collected by the CDF~\cite{cdfref} 
and D$\O$~\cite{d0ref} detectors. In particular, Higgs bosons in supersymmetric models and 
fermiophobic scenario have been investigated, and limits on production cross sections and 
theory parameters have been established.}

\section{Introduction}

The CDF and D$\O$ experiments are finally reaching sensitivity to a standard model
Higgs boson production in $p\bar p$ collisions at the Tevatron~\cite{SMLimits}.
Nevertheless, no hint for Higgs has been observed yet. Moreover, the experiments
can not still probe the low mass region $M_H<160$~GeV/c$^2$ which is favorite by
the fit to the electroweak observables. 

Searches for Higgs boson production in the context of beyond standard model theories
are then well motivated and have been carried out both from CDF and D$\O$ collaborations.
We will summarize here the latest results, by focusing on four different scenarios:
neutral Higgs bosons in the minimal supersymmetric standard model (MSSM), charged Higgs
bosons, Higgs in the next to minimal supersymmetric standard model (nMSSM), and
fermiophobic Higgs bosons. 

\section{Neutral Higgs Bosons in the MSSM}

The MSSM requires the existence of two isodoublets of Higgs fields, which couple 
to up-type and down-type fermions respectively. Out of the eight degrees of freedom,
three are absorbed by the masses of the $Z$ and $W$ bosons, and five are associated to 
new scalar particles: three neutral Higgs bosons ($h$, $H$, $A$) and two charged ones
($H^\pm$). At tree level, Higgs phenomenology in the MSSM is described by two 
parameters: the ratio $\tan\beta$ of the vacuum expectation values of the Higgs
doublets, and the mass $m_A$ of the pseudoscalar boson $A$. 

The couplings of neutral Higgs bosons to bottom quark $b$ and tau $\tau$ (down-type
fermions) scale as $\tan\beta$ with respect to standard model value. For $\tan\beta\sim 1$,
therefore, limits on standard model Higgs production apply to neutral Higgs in MSSM too.
At high values of $\tan\beta$, production processes involving $b$ quarks are enhanced of
a factor $\tan^2\beta$. Moreover, the pseudoscalar boson $A$ becomes degenerate with 
either one of the other neutral Higgs particles, which provides a further enhancement of
the searched signal. Finally, in the high $\tan\beta$ region the neutral Higgs bosons decay 
dominantly into $b\bar b$ (Br$\sim 90\%$) or $\tau^+\tau^-$ (Br$\sim 5$-$13\%$) pairs.

\begin{figure}[t]
\psfig{figure=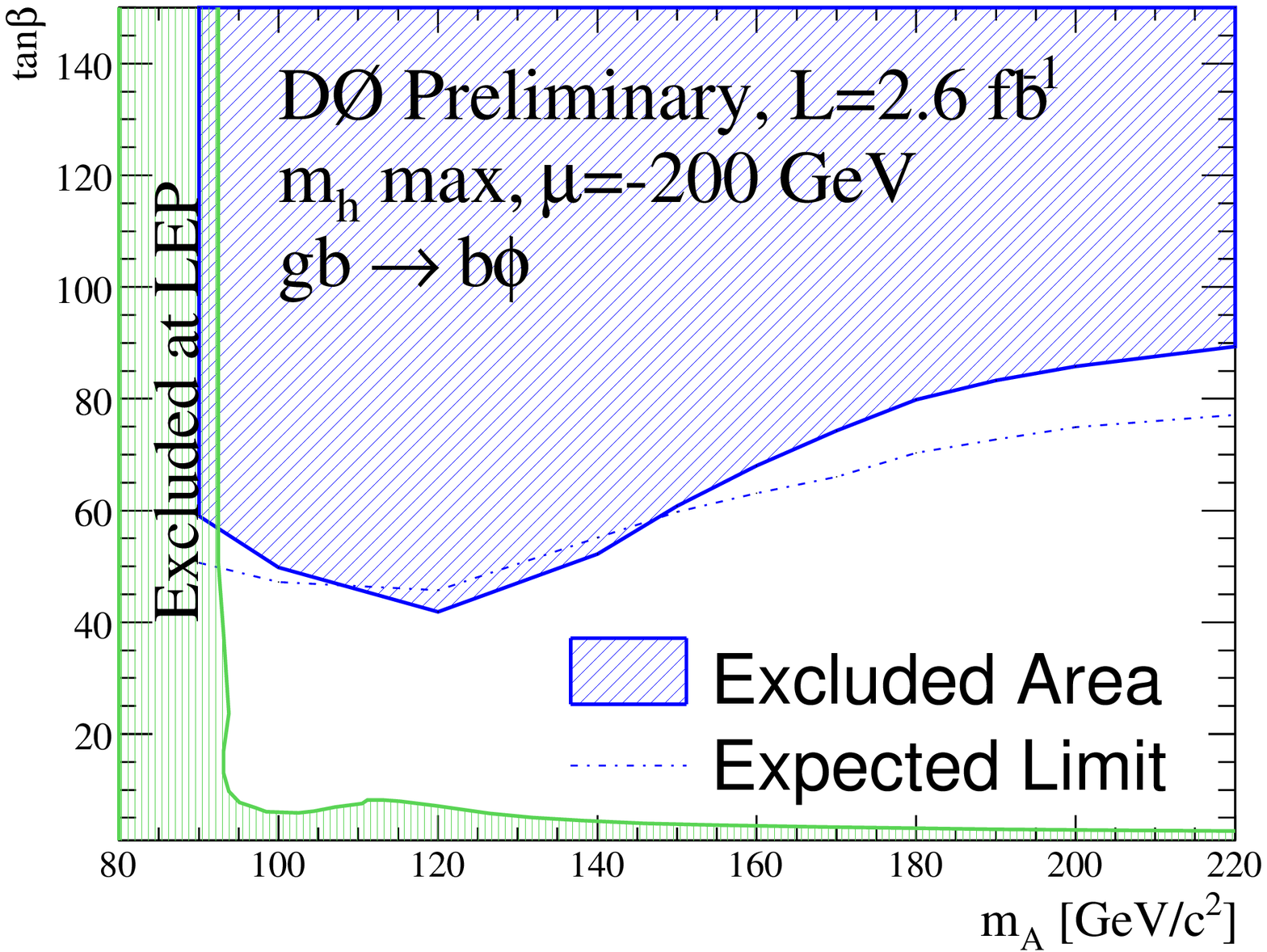,height=61mm}
\psfig{figure=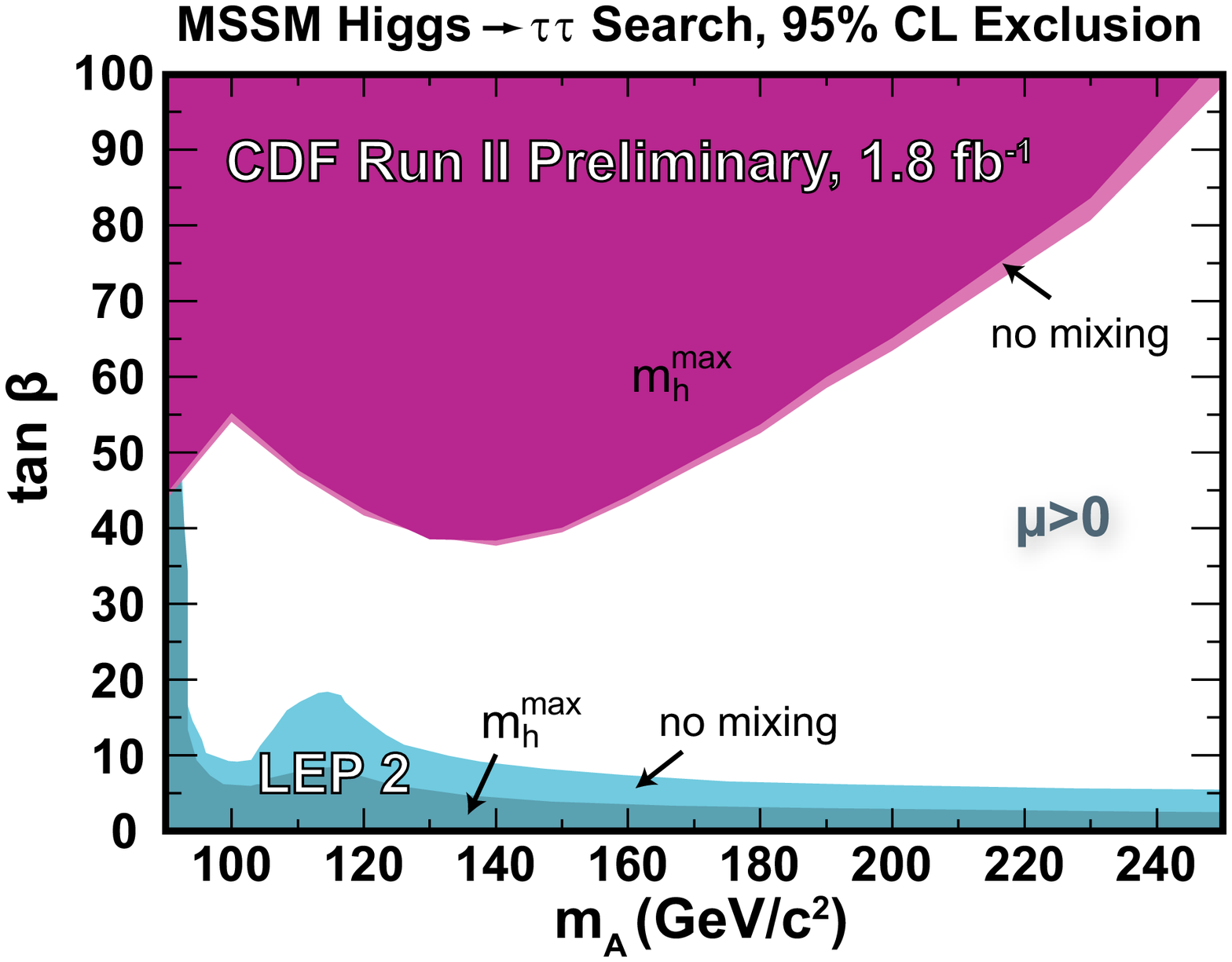,height=61mm}
\caption{Exclusion regions in the plane $m_A$-$\tan\beta$ from neutral Higgs 
boson $\phi^0$ searches at the Tevatron. Left plot result comes from $\phi^0b\rightarrow bbb$
studies at the D$\O$ experiment, while right one has been obtained by the CDF collaboration 
by searching for inclusive $\phi^0\rightarrow \tau^+\tau^-$ production.}
\label{f:mssm} 
\end{figure}

The CDF and D$\O$ collaborations looked for signal of MSSM neutral Higgs boson production both 
inclusively and in association with a bottom quark. While offering a higher cross section,
the inclusive production can only be exploited in the decay mode to taus, due to the high
background from QCD processes which can mimic a $b\bar b$ signal. Associated production has
been instead investigated both in the $\tau^+\tau^-$ and $b\bar b$ decay channels. The
reconstruction of the hadronic decays of the tau and the identification of jets coming from
b quark hadronization are key ingredient of these searches. Upper limits on production
cross sections can be interpreted as exclusion regions in the plane $m_A$-$\tan\beta$.
Since at higher order other parameters of the MSSM become important for Higgs phenomenology, 
a particular set (benchmark scenario) for their values have to be considered when drawing 
the exclusion regions. Fig.~\ref{f:mssm} shows the results for the maximum Higgs mass and the
no-mixing scenarios~\cite{scenarios}, and for Higgs mixing parameter $\mu=\pm 200$~GeV. 

\section{Charged Higgs Bosons}

\begin{figure}[t]
\psfig{figure=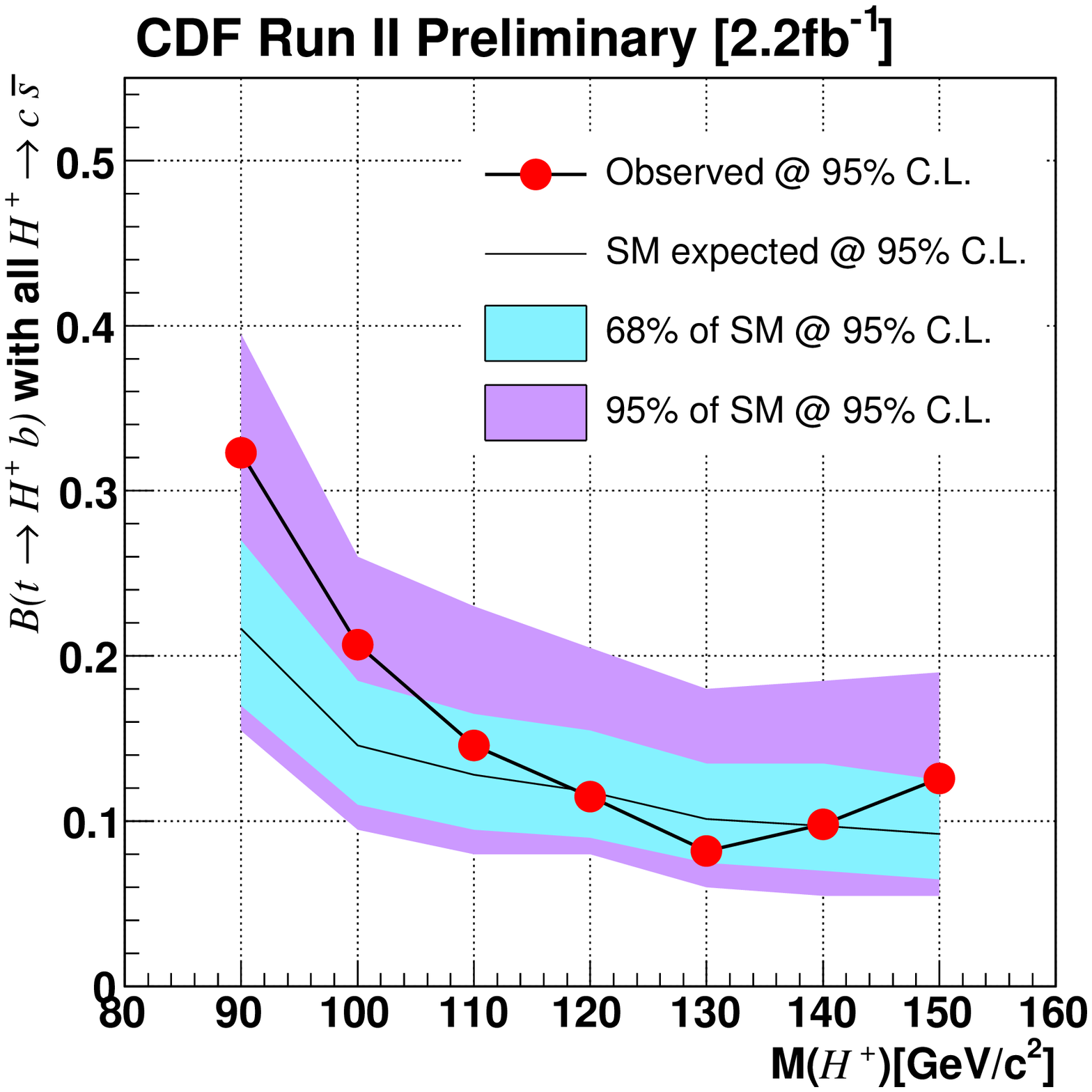,height=61mm}
\psfig{figure=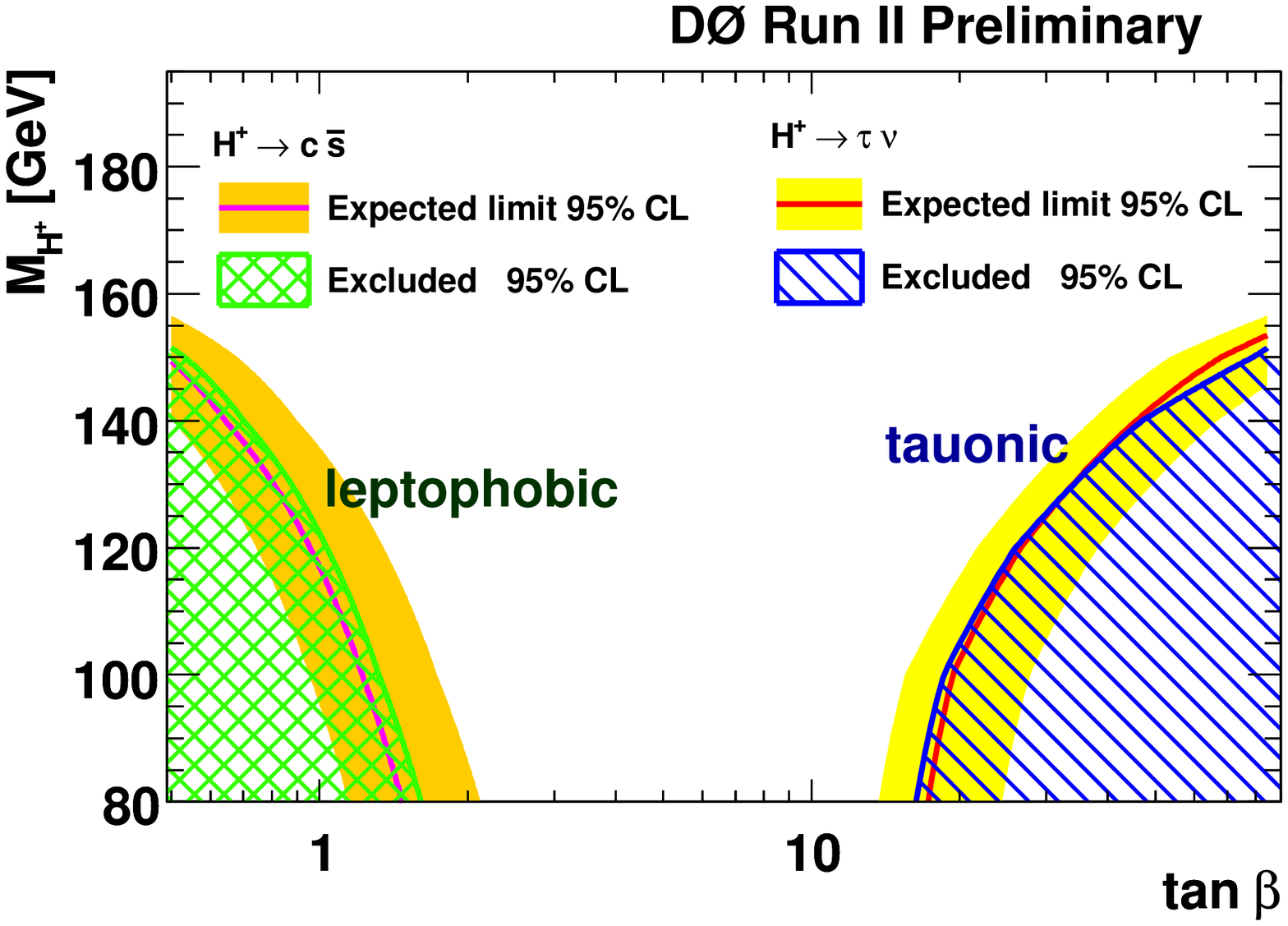,height=61mm}
\begin{minipage}{0.41\linewidth}
\caption{CDF upper limits on Br$(t\rightarrow H^+b)$ when Br$(H^+\rightarrow c\bar s)=100\%$ is assumed.}
\label{f:hcrgcdf} 
\end{minipage}
\hspace{0.02\linewidth}
\begin{minipage}{0.56\linewidth}
\caption{D$\O$ exclusion regions in the plane $\tan\beta$-$M_{H^\pm}$ for a leptophobic and a tauonic model of 
the charged Higgs decay.}
\label{f:hcrgd0} 
\end{minipage}
\end{figure}

Searches for charged Higgs bosons $H^\pm$ have been carried out at the Tevatron experiments by 
looking in top quark samples. In particular, the CDF collaboration looked for the decay of top 
quark into charged Higgs and bottom quark in $t\bar t$ pair production events. In order to reduce 
background, the other top was required to decay in a $W$ boson which then decay to leptons, and 
the bottom quarks are required to be tagged. The charged Higgs is assumed to decay exclusively 
to quarks. This search is sensitive to MSSM production for $\tan\beta< 1$ and 
$M_H^\pm< 130$~GeV/c$^2$. By fitting the observed dijet mass distribution to 
$H^\pm\rightarrow q\bar q^\prime$, $W^\pm \rightarrow q\bar q^\prime$ and background templates,
an upper limit on the branching ratio of the $t\rightarrow H^+b$ decay has been set as a function
of the Higgs boson mass (see Fig.~\ref{f:hcrgcdf}).

The D$\O$ experiment searched for charged Higgs boson by using a different approach, which consists 
in computing the effects that a $t\rightarrow H^+b$ decay would have on the yields of events
in the different $t\bar t$ decay channels, and then comparing the expectations to the observed
number of events to set limit on the branching ratio of top quark decay to charged Higgs boson.
Fig.~\ref{f:hcrgd0} shows the results in two scenarios for the Higgs decay: a tauonic model where
the Higgs decays exclusively into tau and neutrino (which is equivalent to the MSSM for very 
high values of $\tan\beta$), and a leptophobic model assuming Br($H^+\rightarrow c\bar s$)$=100~\%$
(realized by a general multi-Higgs-doublet model~\cite{mhdm}).

\section{Higgs Bosons in the nMSSM}

The nMSSM~\cite{nmssm} adds a singlet superfield to the MSSM, allowing the theory to generate dynamically
the mixing term $\mu H_uH_d$ in the Higgs sector, and solving in this way the $\mu$ problem. It also
turns out to be the simplest supersymmetric model in which the electroweak scale originates only from the 
scale of supersymmetry breaking. 

Two additional Higgs boson states appear in the nMSSM: a neutral CP-even Higgs $s$ and a CP-odd Higgs $a$. 
While the lightest CP-even Higgs boson $h$ remains SM-like in the nMSSM, its dominant decay may not be 
necessarily into a $b\bar b$ pair, since the mass of the new state $a$ is allowed to be small enough for 
the decay $h\rightarrow aa$ to become dominant. LEP limits on the mass of the $h$ boson can then 
be avoided if $M_{a}<2m_b$, obtaining in this way a theory free from fine-tuning problems. 

The D$\O$ collaboration searched for the nMSSM process $h\rightarrow aa$. At low $M_a<2m_\tau$, a 4 muon signature is
required, and upper limits on $\sigma(p\bar p\rightarrow hX)\times $Br$(h\rightarrow aa)\times $Br$(a\rightarrow \mu\mu)^2$ 
at about $10$~fb have been set. Assuming Br$(h\rightarrow aa)\approx 100\%$ and $M_h=120$~GeV/c$^2$, which correspond to a 
production cross section of 1000~fb within the SM, it should be Br$(a\rightarrow\mu^+\mu^-)\apprle 10\%$ to avoid detection,
while the nMSSM predicts a branching ratio for the decay $a\rightarrow\mu^+\mu^-$ greater than $10\%$ for $a$ boson mass
up to $2m_c$, and, depending on the branching ratio of $a$ to charm quarks, possibly even up to $2m_\tau$.
For $M_a>2m_\tau$, the decay channel to $\mu^+\mu^-\tau^+\tau^-$ has been investigated and the limits set on Higgs production
are still a factor of $\sim 4$ larger than predictions.

\section{Fermiophobic Higgs Bosons} 

A fermiophobic Higgs boson would greatly enhance the sensitivity of the Tevatron 
experiments to Higgs production in the low mass region ($M_H\apprle 130$~GeV/c$^2$), where the
dominant SM decay to $b\bar b$ provides a difficult signature due to the background from
QCD processes. Theoretically, null (or highly suppressed) coupling of the Higgs boson to
fermions could indicate a different origin for fermion and boson masses. 

The benchmark fermiophobic model assumes the same Higgs couplings to gauge boson as in
the SM, and no couplings to fermions. In such a scenario, Higgs direct production is forbidden,
and productions in association with a $W$ or a $Z$ boson and via vector boson fusion become 
the dominant mechanisms. 

\begin{figure}[t]
\psfig{figure=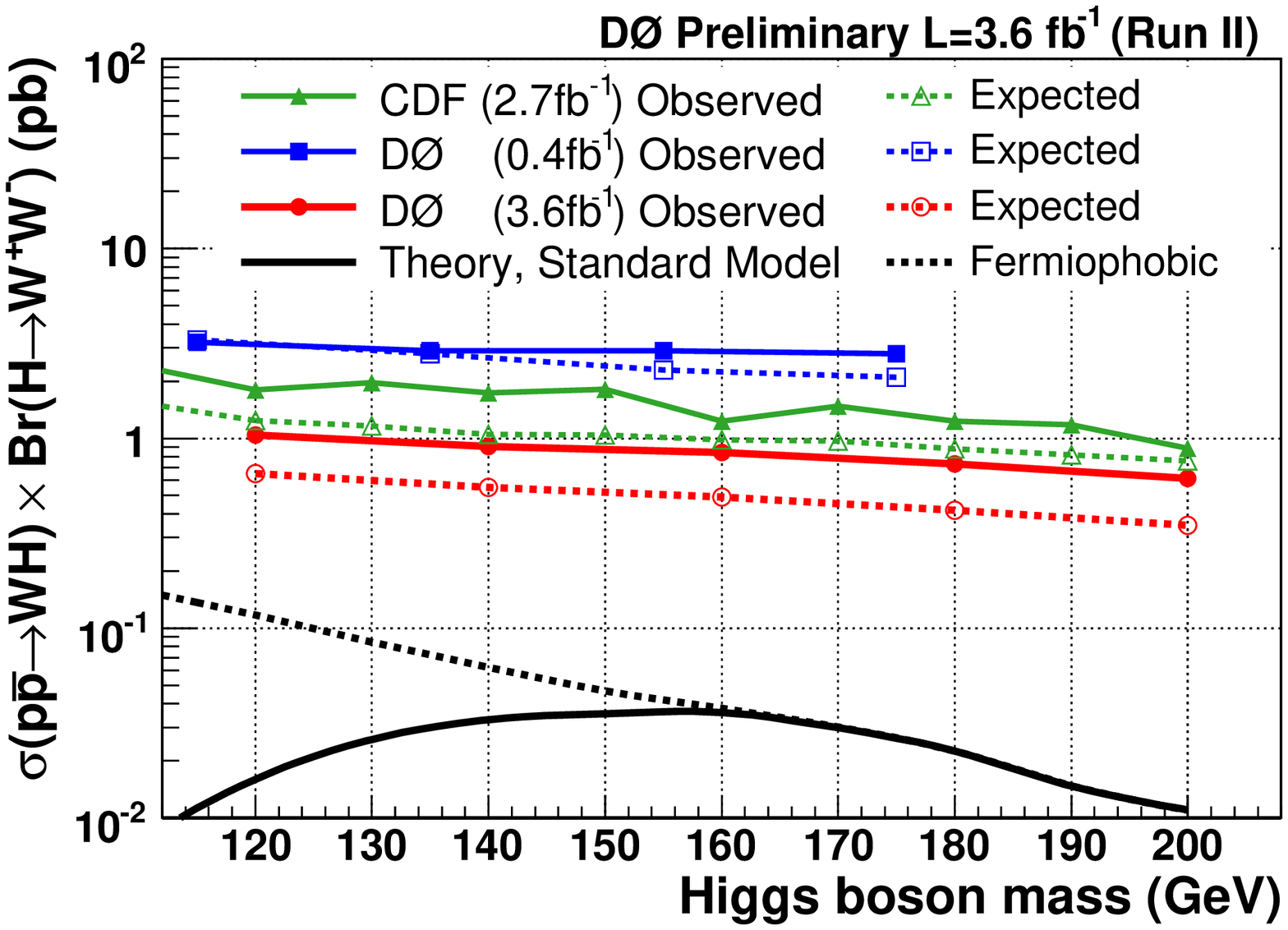,height=61mm}
\psfig{figure=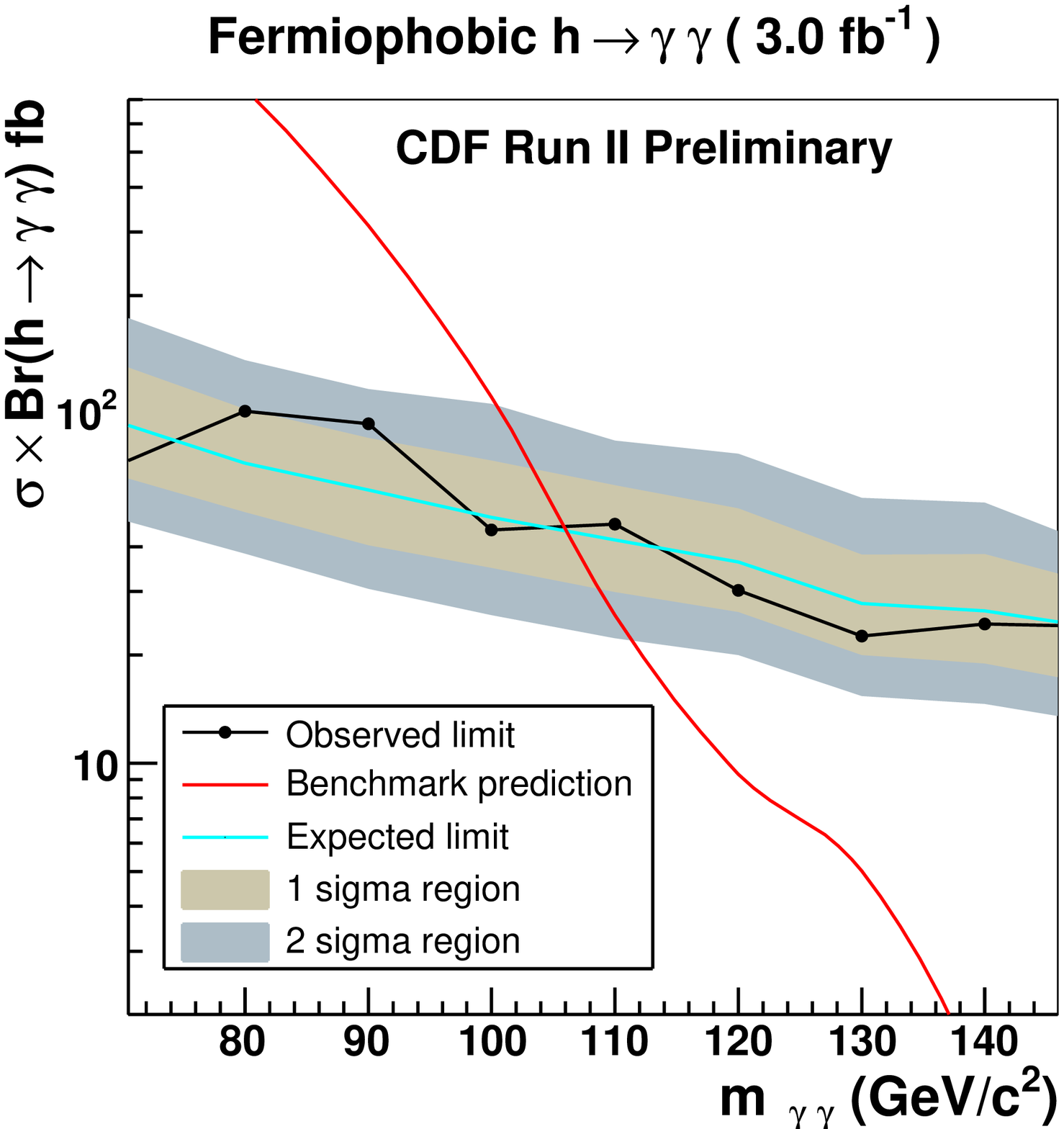,height=65mm}
\begin{minipage}{0.55\linewidth}
\caption{CDF and D$\O$ upper limits on the Higgs associated production with a $W$ boson times the branching ratio 
of Higgs decays to $W^+W^-$. Also shown are the predictions from the standard model and the benchmark fermiophobic
model.}
\label{f:whwww} 
\end{minipage}
\hspace{0.02\linewidth}
\begin{minipage}{0.42\linewidth}
\caption{CDF upper limits on the production cross section times branching ratio for a fermiophobic Higgs boson 
decaying into photons, compared with benchmark model predictions.}
\label{f:hff} 
\end{minipage}
\end{figure}

The CDF and D$\O$ collaborations looked for $WH\rightarrow WWW^*$ production in events with
two leptons (electrons or muons) with the same charge. Observed limits on the production 
cross section times the branching ratio for the decay $H\rightarrow W^+W^-$ are compared
to SM and fermiophobic model predictions in Fig.~\ref{f:whwww}. 

Inclusive production of a Higgs boson decaying to photons has also been searched by the two
experiments by exploiting the high resolution (about $3\%$) on the reconstructed mass of the 
diphoton system provided by their calorimeters. When comparing the observed limits on the 
production cross section to the benchmark model expectations, a lower limit on the mass 
of a fermiophobic Higgs boson is set at $106$~GeV/c$^2$ (see Fig.~\ref{f:hff}). 

\section{Conclusions}

The CDF and D$\O$ collaborations looked actively for Higgs bosons in the context of physics
beyond the standard model in about 1-4~fb$^{-1}$ of $p\bar p$ collisions at the Tevatron collider. 
Advanced techniques have been established and several limits on relevant parameters for different 
theories have been set, but no Higgs production signal has been observed yet. 

Lot of improvements are to come: increased statistics (both experiments already have 5~fb$^{-1}$ 
of data on tape) and combination of different search channels and experiment results will enhance 
the sensitivity to Higgs production, eventually leading to new insights on the mechanism of electroweak
symmetry breaking.


\end{document}